\documentclass[a4paper,twoside]{article}

\usepackage{epsfig}
\usepackage{url}
\usepackage{subcaption}
\usepackage{calc}
\usepackage{amssymb}
\usepackage{amstext}
\usepackage{amsmath}
\usepackage{amsthm}
\usepackage{multicol}
\usepackage{xcolor}
\usepackage{pslatex}
\usepackage{apalike}
\usepackage[bottom]{footmisc}
\usepackage{SCITEPRESS}     

\begin{document}

\title{A Novel OMNeT++-based Simulation Tool for Vehicular Cloud Computing in ETSI MEC-compliant 5G Environments}

\author{\authorname{Angelo Feraudo\sup{1}, Alessandro Calvio\sup{1}, and Paolo Bellavista\sup{1}}
\affiliation{\sup{1}Department of Computer Science and Engineering, University of Bologna, Viale Risorgimento 2, Bologna, Italy}
\email{\{name.surname\}@unibo.it}
}

\abstract{Vehicular cloud computing is gaining popularity thanks to the rapid advancements in next generation wireless communication networks. Similarly, Edge Computing, along with its standard proposals such as European Telecommunications Standards Institute (ETSI) Multi-access Edge Computing (MEC), will play a vital role in these scenarios, by enabling the execution of cloud-based services at the edge of the network.  Together, these solutions have the potential to create real micro-datacenters at the network edge, favoring several benefits like minimal latency, real-time data processing, and data locality. However, the research community has not yet the opportunity to use integrated simulation frameworks for the easy testing of applications that exploit both the vehicular cloud paradigm and MEC-compliant 5G deployment environments. In this paper, we present our simulation tool as a platform for researchers and engineers to design, test, and enhance applications utilizing the concepts of vehicular and edge cloud. Our platform significantly extends OMNet++ and Simu5G, and implements our ETSI MEC-compliant architecture that leverages resources provided by far-edge nodes. In addition, the paper analyzes and reports performance results for our simulation platform, as well as provides a use case where our simulator is used to support the design, test, and validation of an algorithm to distribute MEC application components on vehicular cloud resources.}


\onecolumn \maketitle \normalsize \setcounter{footnote}{0} \vfill
\section{\uppercase{Introduction}}
Next generation wireless communication networks are advancing at a rapid pace, leading to the development and prototyping of highly innovative application scenarios. 
In this context, as the amount of connected vehicles rapidly increases ~\cite{elektrobit}, drivers can take advantage of a wide range of distributed applications and services, very often cloud-based, as in other vertical domains,  including improved access to information and entertainment features. However, it starts to be widely recognized that traditional client-to-cloud architectures cannot meet the stringent requirements of many time-sensitive Internet of Things (IoT) applications. Furthermore, with the constant growth of intelligent cars, equipped with a large set of onboard sensors \cite{sabellafullyconnected}, the amount of vehicle-generated data may pose a significant challenge to the backbone network.

Edge Computing has emerged as a promising solution for such emerging distributed environments, as it offers a more efficient and effective way to handle the ever-increasing demand for computing and storage resources. Moreover, edge computing has the potential to replace traditional cloud solutions due to its potential to reduce network stress, decrease latency, improve efficiency, and enable real-time data processing. To accelerate the adoption of this paradigm, the European Telecommunications Standards Institute (ETSI) has proposed the Multi-access Edge Computing (MEC) standard~\cite{etsiwebsiterefarch} for cloud micro-datacenters situated at the edge of Radio Access Network (RAN). The MEC architecture offers a virtualized infrastructure that orchestrates, distributes, and manages the life-cycle of MEC-compliant applications enabling their execution as close as one step from the involved datasources and/or users. In this way, MEC aims at becoming a key technology for emerging radio networks (e.g., 5G), as it empowers legacy radio units with cloud computing capabilities~\cite{7931566}. 

Recently, Vehicular Cloud Computing (or more simply Vehicular Computing) has been proposed as a promising paradigm to support distributed applications designed according to the edge cloud approach ~\cite{6257116,olariu2011taking}. Vehicular Computing exploits the computing power locally available on today's vehicles to create cost-effective mobile clouds at the far-edge layer. The formation of these dynamic clouds should be achieved autonomously by vehicles, which share their resources among them and/or with nearby MEC nodes to extend their virtualized resources for service execution. 

Numerous research efforts have been devoted to developing architectures that leverage the resources dynamically available on vehicles~\cite{8936985}. To enhance resource availability at the network edge, various works have proposed to exploit the underutilized computational power of both stationary~\cite{5935198,8463481} and moving vehicles~\cite{7415983}. These opportunistic resources can be utilized for diverse purposes to handle the growing number of applications used in vehicular networks. For instance, vehicles can serve as relay nodes~\cite{5935198} to improve network connectivity, or as computing nodes~\cite{8522034,9123902,9344808} to reduce the impact of these applications on the performance of edge nodes.

Moreover, some recentstudies have focused on providing simulation frameworks~\cite{ahmed2019services} since practical experiments on vehicular network environments are expensive and challenging. In fact, to validate their proposals, some of the works in the literature ~\cite{9366768,9344808,9123902,7946184} have utilized these simulation frameworks mainly for i) generating vehicle traces and ii) simulating the behavior of vehicular network protocols. To the best of our knowledge, there is currently no simulation tool that offers a single vehicular computing-based platform where researchers can design/test their algorithms and applications while exploiting at the same time the vehicular cloud computing paradigm and the MEC standard deployment environment.

In this paper, we propose a simulation tool that leverages the ETSI MEC standard to incorporate the resources available at the edge of the network, such as those provided by collaborating vehicles, into the cloud continuum spectrum. The simulation tool implements an extended version of the ETSI standard ~\cite{feraudo2023novel}, which enhances MEC-compliant edge nodes with resources from vehicles within a designated Area of Interest (AoI). The resources provided by the nodes are registered in the edge resource pool and can be accessed through standardized interfaces. To accommodate node mobility, the extended MEC architecture supported by our simulation platform  assists devices during application migrations by triggering the user context transfer to MEC applications running on nodes leaving the AoI through standard defined API~\cite{etsiamsapi}. As an extension
of the ETSI MEC standard, the modeled architecture allows dealing with some of the primary challenges arising in vehicular computing environments, such as integration with cloud resources and enabling coexistence among heterogeneous technologies. Furthermore, it allows dealing with resource volatility issues (i.e., nodes that dynamically join/leave during service provisioning) via a standardized migration mechanism. In particular, to build our simulation platform, we utilized the OMNeT++ network simulator as the underlying framework and incorporated the Simu5G library to model the 5G network and communications aspects; more details about these tools will be provided in the next sections. 

The remainder of the paper is structured as follows. Section~\ref{sec:background}  provides an overview of the related background, which includes a description of our previous preliminary work on which basis we have developed our simulation platform. Section~\ref{sec:impl} presents the interactions and modules that we have originally implemented to create a MEC-compliant vehicular computing environment in a 5G network. We then evaluate the performance of our simulation platform while performing resource management and by providing a practical example of its usage in section~\ref{sec:perf}.




\begin{table}
\centering
\caption{Table of acronyms for MEC elements}
\resizebox{\linewidth}{!}{{\renewcommand{\arraystretch}{1.3}
\begin{tabular}{c c}
\hline
 \textbf{Abbreviation} & \textbf{Definition} \\
 \hline
 AMS & Application Mobility Service \\
 MEC & Multi-access Edge Computing \\
 MEC-App & MEC Application \\
 MEC-H & MEC Host \\
 MEC-O & MEC Orchestrator \\
 MEC-P & MEC Platform \\
 MEC-PM & MEC Platform Manager \\
 UALCMP & User Application LifeCycle Management Proxy \\
 VI & Virtualisation Infrastructure \\
 VIM & Virtualisation Infrastructure Manager \\
 \hline
\end{tabular}}}
\label{table:mec_acr}
\end{table}
\section{\uppercase{Background}}\label{sec:background}
\subsection{ETSI Multi-access Edge Computing}\label{subsec:MEC}

\begin{figure*}[ht]

    \centering
    \includegraphics[width=1\linewidth]{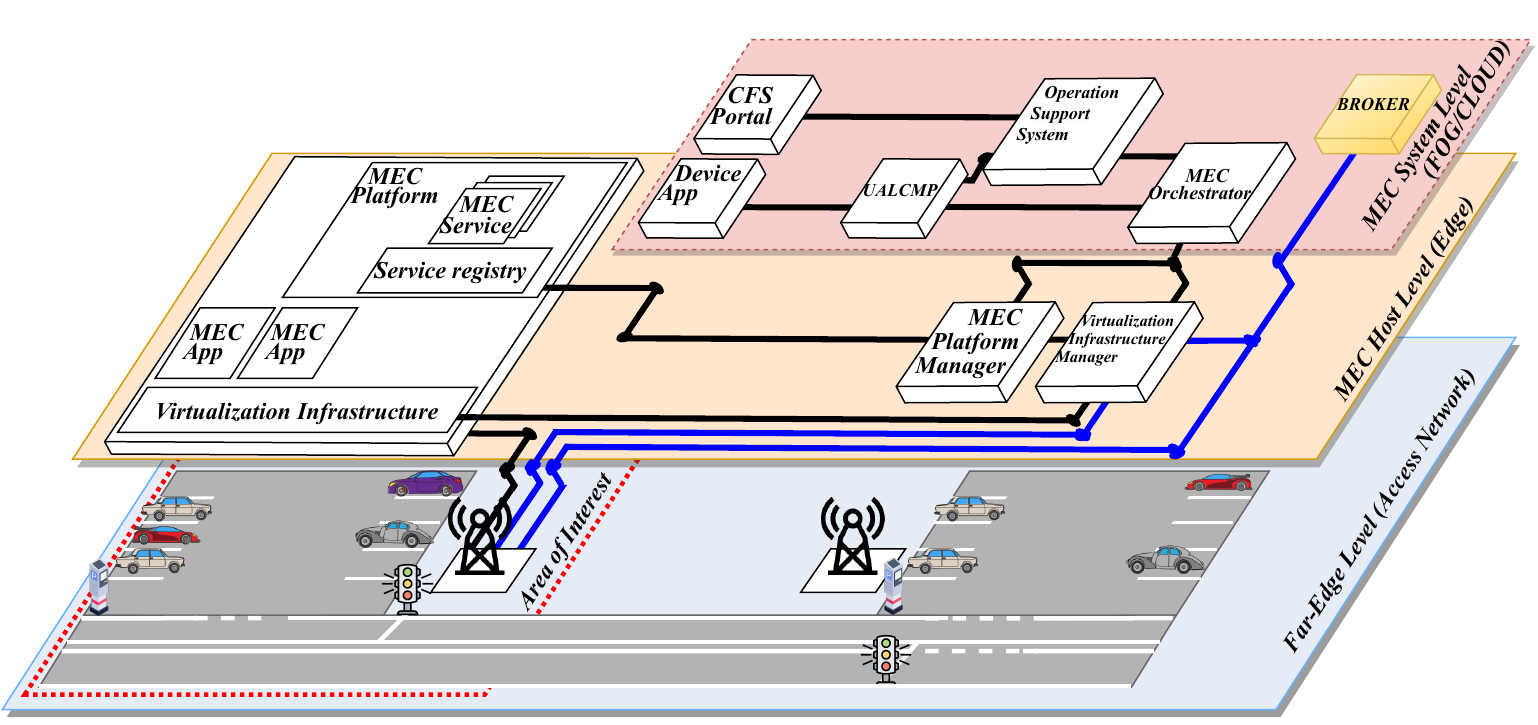}
    \caption{The extended MEC architecture, already presented in \cite{feraudo2023novel}}
    \label{fig:arch}

\end{figure*}

Figure~\ref{fig:arch} illustrates the MEC reference architecture proposed by ETSI in~\cite{etsiwebsiterefarch}. It offers the opportunity to run contextualized MEC-compliant applications within a virtualized and multi-tenant environment. The reference architecture consists of two layers: the system level and the host level, which together comprise all standard functional elements. The system level works as an entry point for authorized users to request the execution of MEC applications. It includes management-related functional elements, such as \textit{MEC Orchestrator} (MEC-O) and \textit{User Application LifeCycle Management Proxy} (UALCMP), as well as other components that are necessary for standard operation, e.g.,\textit{ Device Application}. The MEC-O is the core functionality for layer management, as it has a global view of the MEC-compliant edge nodes (\textit{MEC Hosts}) present in a particular area and serves as a controller for MEC application distributions. By selecting the best MEC Host based on the requirements of the requested services, the MEC-O triggers the instantiation and termination of applications. The second management element, i.e., the UALCMP, acts as an intermediary node between the user and the application. It supports instantiation and termination requests from the users, which are forwarded to the MEC-O. \textit{Device application} operates on user devices and interacts with the MEC system through the UALCMP.

The host level is situated at the edge of the network and is responsible for providing and managing the virtualization platform where MEC applications are deployed. It comprises the edge node that provides virtualized computational, network, and storage resources, i.e., MEC-H. The MEC-H enables MEC application deployment through the \textit{Virtualization Infrastructure} (VI), which is also responsible for managing the data plane among the network interfaces through traffic policies. The MEC-H also includes a \textit{MEC Platform} (MEC-P) that exposes a service registry, thus enabling the supported applications to discover, offer, and consume standard services. The defined standards include Radio Network Information Service, Location Service, and Application Mobility Service, among the others. Moreover, at this level, the \textit{Virtualization Infrastructure Manager} (VIM) and the MEC Platform Manager (MEC-PM) act as the management-related functional elements and serve as access points for the MEC-O to the lower layer. The VIM is responsible for managing and releasing the virtualized resources and configuring the VI to run software images appropriately. 


In the recent document released by the ETSI group~\cite{etsiwebsiterefarch}, the MEC reference architecture has been updated with an additional functional element, the MEC Federator. This element is especially noteworthy due to the increased recognition that the standard has received, and it will serve as the core element for federating multiple MEC systems that belong to different network domains.

\subsection{Vehicular Computing Paradigm}
Many current and future applications for connected vehicles can take advantage of edge resources, including infotainment and time-sensitive applications related to traffic safety. The integration of vehicular networks with the MEC infrastructure can be referred as Vehicular Edge Computing (VEC) ~\cite{liu2021vehicular}. VEC aims to bring computational capabilities to the proximity of vehicular users, allowing for services to be readily available via Vehicle-to-Infrastructure (V2I) communications. Nevertheless, the ever-increasing number of applications and businesses is starting to make the infrastructure unable to efficiently satisfy their demands and possibly stringent requirements, e.g., on latency. For this reason, the need has emerged to identify new resources that can support the more traditional edge infrastructure.

In \cite{olariu2011taking} and \cite{6257116}, the authors introduced the Vehicular Cloud Computing concept relying on the idea that modern-day vehicles come equipped with powerful on-board computers, ample storage, and an array of sensing devices. In their work \cite{olariu2011taking}, the authors define Vehicular Computing as a collaborative way to share resources among vehicles to solve problems that would otherwise require a significant amount of time with a more traditional centralized architecture, in particular for context-specific applications. In line with this, in~\cite{6257116} the author states that the vehicular cloud paradigm allows keeping the information gathered by vehicle sensors locally and sharing it solely with other vehicles, as the sheer volume of in-vehicle generated data can pose serious technical challenges for the network infrastructure. According to these definitions, a cloud of vehicles, i.e., a set of vehicles that serve as computation nodes for a diverse range of services, can be formed anywhere on the road and their onboard resources can be dynamically allocated to authorized users. Despite its potential benefits, such a vehicular computing environment also poses several challenges that must be addressed. These challenges include distributed ownership, as each vehicle has a single owner responsible for deciding whether to share onboard resources; high node mobility, which makes it difficult to predict the vehicular residency times in the cloud even when clouds are formed using resources of vehicles within a parking lot; device heterogeneity, as vehicles are manufactured by different companies; security and privacy.

To deal with some of the aforementioned challenges, we recently presented an ETSI MEC-compliant architecture that enlarges the edge resource pool through vehicular computational resources ~\cite{feraudo2023novel}. In the next section, we provide additional information about this architecture, which serves as the architectural basis for the simulation platform originally described in this paper. 

\subsection{Using Vehicle Resources to Expand MEC Nodes}\label{sec:pwork}
In our previous work~\cite{feraudo2023novel}, we consider the evidence that present-day far-edge nodes, e.g., vehicles, hold sufficient computing and storage resources to form micro-datacenters at the network edge. The proposal exploits the edge node components of the ETSI MEC standard, i.e., MEC-H, to leverage far-edge node resources as part of its resource pool. Hence, compared to the traditional ETSI MEC architecture, the proposed extension encompasses mechanisms capable of deploying and distributing applications on local and remote resources by keeping an eye on resource volatility issues.

 As illustrated in figure~\ref{fig:arch} (blue lines), each MEC-H defines an Area of Interest (AoI) corresponding to the area within which far-edge device resources are collected. The AoI may coincide with the base station coverage, thus it depends on where the MEC-H is located, i.e., at either the network edge (close to the base station) or the central data network (at aggregation points). The latter location enables MEC-Hs to subscribe for multiple AoI, as subscriptions may depend on the latency introduced by the network. Indeed, for MEC-Hs physically collocated with the base station, subscribing to several AoIs may result in huge costs in terms of latency, as the distance with other base stations increases.

 To model the resource acquisition procedure, our extended MEC architecture introduces an external entity at the MEC system level running a \textit{Broker}, which represents the message broker of a publish-subscribe system. It allows MEC-H subscriptions to the AoI and manages their notification whenever a new device enters or leaves the area. The same entity runs a reward system encouraging far-edge devices to lease their local virtualized resources, e.g., computing power, and join the resource pool. It relies on a device-initiated scheme requiring far-edge nodes to request available rewards contextualized to the AoI. Thus, whenever a new device accepts the rewards indicated by the MEC-H related to that area, it publishes the amount of resources that it is willing to make available. In addition, once a device leaves the AoI, the MEC-H receives the notification, removes the concerned resources from those available in the pool, and starts the mobility procedure for the apps running on that device.

 By delving into slightly finer technical details, the VIM is the main MEC-H internal entity to be affected during resource acquisition, as it is in charge of administering the MEC-H resource pool and preparing the VI for the deployment of MEC applications. Indeed, in our MEC extended architecture ~\cite{feraudo2023novel}, the VIM operates to handle a heterogeneous pool of distributed resources. Specifically, once registered to the MEC-O, the VIM specifies the content of interest to the \textit{Broker} (content-based subscription) corresponding to the AoI parameters (e.g., circle center and diameter) given during its configuration. Thus, when the \textit{Broker} notifies it of new device resource acquisition, the VIM stores the endpoint information corresponding to an external VI address and the resource capacity of that device. This allows the VIM to keep track of the single contributions that each host brings in terms of computational resources. In this context, it is beneficial to differentiate between infrastructure resources and transient ones that are added based on the availability of far-edge nodes in the vicinity.

 This section overviewed the architecture that underlies the simulation platform originally presented in this paper. Let us point out that, while the presented architecture is designed to be versatile and to support different types of edge nodes offering virtualized resources at the far-edge layer, the current implementation of our simulation platform supports the exploitation of only parked vehicles as far-edge nodes.



\begin{figure*}[ht]

    \centering
    \includegraphics[width=1\linewidth]{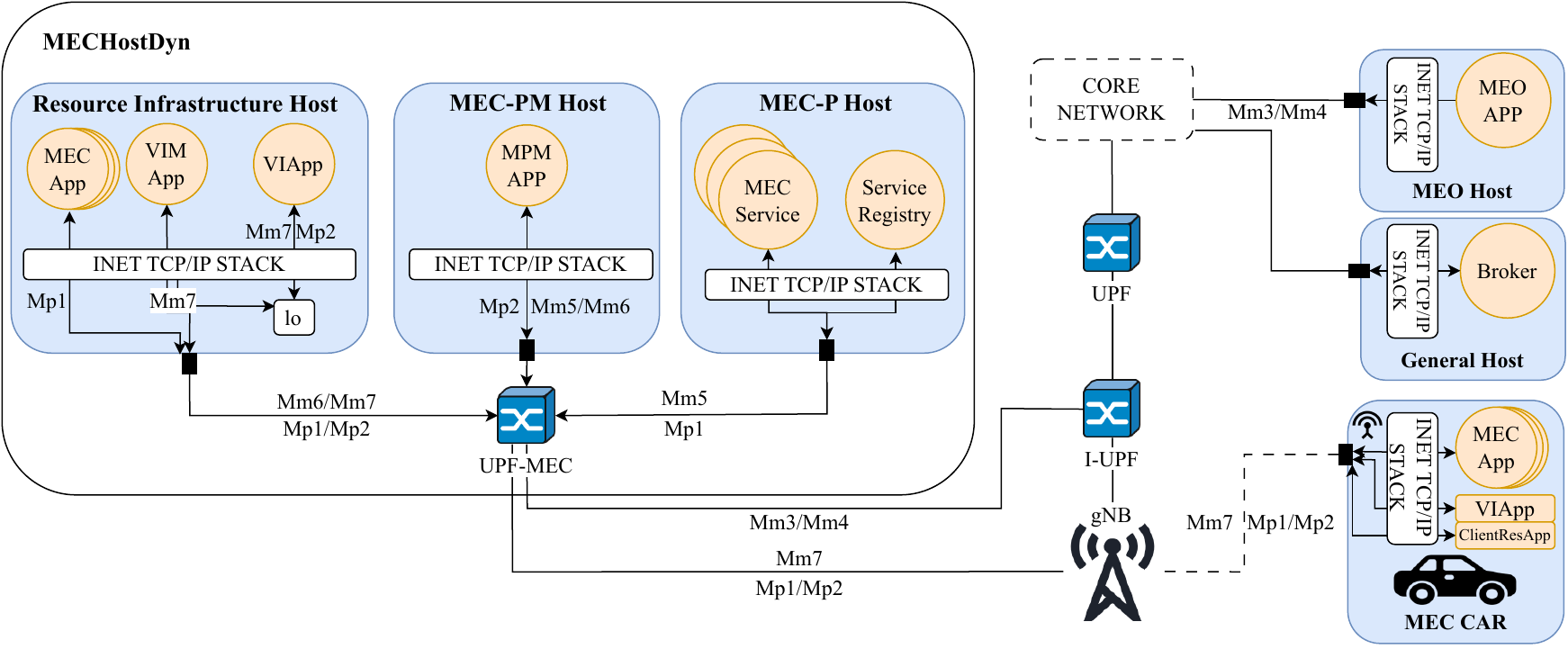}
    \caption{Simulation Tool Modules Structure}
    \label{fig:modules}

\end{figure*}

\section{\uppercase{Our Simulation Platform for MEC-compliant Vehicular Computing}}\label{sec:impl}
The Primary objective of our original simulation tool is to provide a platform for researchers and engineers to design, test, and enhance applications by utilizing the concept of vehicular computing in a 5G environment. The proposed framework works within the OMNeT++ event-based simulator and uses the well-known Simu5G communication library~\cite{nardini2020simu5g}. Specifically, we developed the OMNeT++ modules inside the Simu5G project and reused some of the ETSI MEC modules provided by this library.

In the following sections, we will examine the primary internal components of our simulation platform and how they interact to enable the exploitation of external resource infrastructure in MEC-compliant environments.

\begin{figure*}[ht]
    \begin{subfigure}[b]{0.49\textwidth}
        \centering
        \includegraphics[width=1\linewidth]{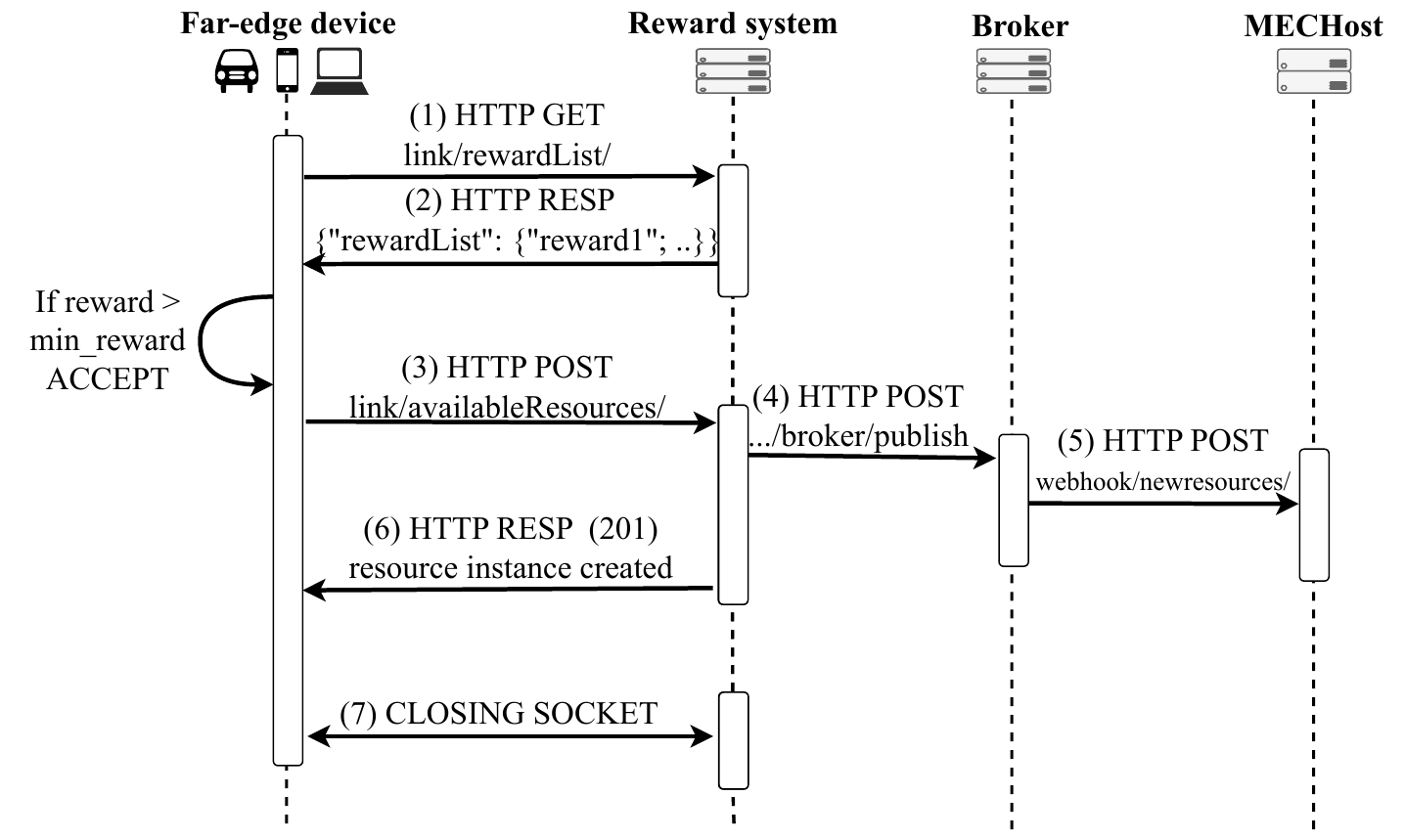}
        \caption{Resource Acquisition Sequence Diagram}
        \label{fig:resSub}
    \end{subfigure}
    \begin{subfigure}[b]{0.5\textwidth}
        \centering
        \includegraphics[width=0.95\linewidth]{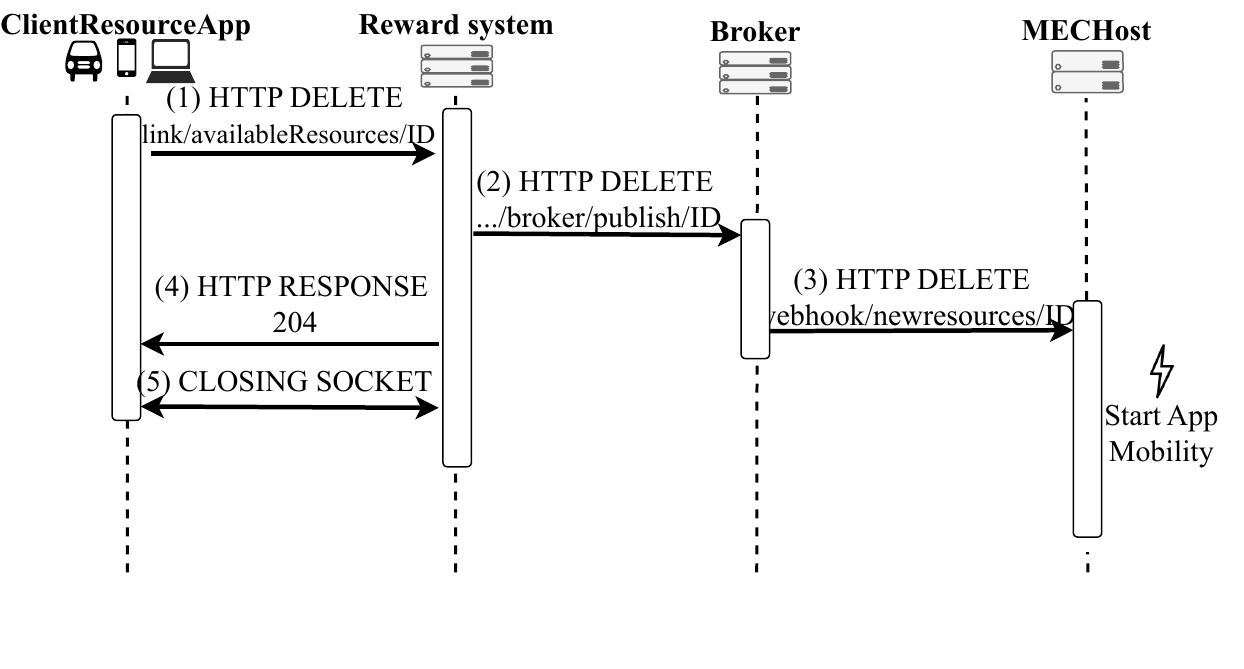}
        \caption{Resource Releasing Sequence Diagram}
        \label{fig:resRel}
    \end{subfigure}
    \caption{Sequence Diagram device-initiated scheme}
    \label{fig:diagramResAl}
\end{figure*}

\subsection{Simulation Tool: Interactions}\label{subsec:interaction}
This sub/section describes the procedures supported for MEC-enabled resource management for vehicular cloud computing, identified on the basis of the MEC extended architecture presented in Section 2. Each procedure requires a specific set of interactions to be completed effectively. These interactions involve both MEC standard and non-standard components, which must be carefully considered to ensure that the necessary resources are acquired, released, and allocated.

\paragraph{Resource Acquisition}
Figure~\ref{fig:resSub} illustrates the steps required by parking vehicles to partake in the resource acquisition procedure. According to the sequence diagram, whenever a new vehicle accepts the rewards indicated by the MEC-H related to that area, it publishes the amount of resources that wants to make available. In addition to resource availability information, the POST request, step (3) in the figure, contains data related to its location, thus allowing the \textit{Broker} to notify the appropriate MEC-H.

\begin{figure*}[ht]

    \centering
    \includegraphics[width=0.8\linewidth]{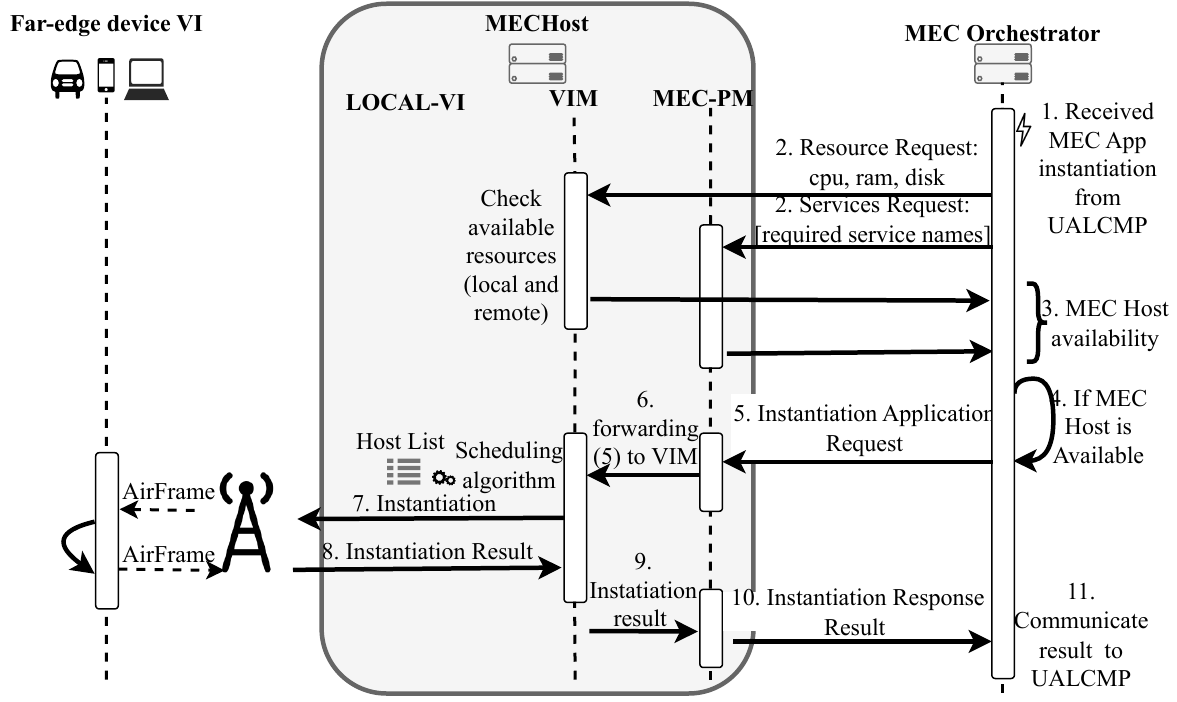}
    \caption{Resource Allocation Sequence Diagram}
    \label{fig:rra}

\end{figure*}

\paragraph{Resource Allocation}
An example of interaction scheme for app instantiation on remote resources is reported in figure~\ref{fig:rra}. The MEC-O receives an instantiation request from the UALCMP and starts resource/service discovery throughout all the MEC-Hs under its management. Once the MEC-H has been chosen, the VIM receives an \textit{instantiation request} (step (7) in Fig.\ref{fig:rra}) and starts the scheduling procedure. As also described in section~\ref{sec:pwork}, when the \textit{Broker} notifies the VIM for new device resource acquisition, the VIM stores the endpoint information, corresponding to an external VI address (see Fig. \ref{fig:modules}), and the resource capacity of that device. Hence, the VIM is aware of the single contributions that each device brings to correctly allocate its resources when needed. The scheduling phase leads to the identification of a single resource infrastructure, either local or remote, where to deploy the requested MEC applications.

\paragraph{Resource Releasing}
Figure \ref{fig:resRel} shows the interactions needed when devices leave the resource pool. In such a scenario, once the corresponding MEC-H receives the notification, it removes the concerned resources from those available in the pool and starts the mobility procedure for the apps running on that device. The migration event exploits the standard Application Mobility Service (AMS) API~\cite{etsiamsapi} defined by ETSI, which currently supports app migrations in environments encompassing multiple edge nodes. Nonetheless, the proposed design in~\cite{feraudo2023novel} requires MEC components to support intra-host migration, e.g., MEC applications mobility from remote to local resource infrastructure. In our simulation platform, we decided to deploy a MEC-assisted application mobility for intra-host migrations, which implies MEC components to trigger the user-context transfer. Thus, once the VIM receives a notification involving nodes leaving the resource pool (step (3) Fig.~\ref{fig:resRel}), it looks for MEC applications running on that host, and creates a migration event for each of them. This series of events serve as inputs for an extended version of the Application Mobility Service, which creates an event chain capable of initiating the intra-host migration of the MEC applications.

\subsection{Simulation Tool: Modules}
In the previous sub-section, we emphasized the involved MEC entities and their interactions when supporting the usage of remote resources provided by collaborating vehicles. To enable these resources to host MEC applications in a simulated environment, we built each of the MEC entities as an application, abstracting their workload from the underlying host. Specifically, to make the extended MEC architecture more agile and flexible, our simulation platform models MEC components as \textit{Inet} Application. 

Figure~\ref{fig:modules} illustrates the deployment of simulation modules that replicate the scenario presented in our proposal (Fig.~\ref{fig:arch}). As already mentioned, our simulation tool currently considers cars as far-edge devices capable of hosting MEC applications. To support this functionality, we introduce the \textit{car} module, which extends the New Radio User Equipment (NRUE) defined in Simu5G. The \textit{car} module acts as a wrapper for any 5G-enabled device, providing computational resources (e.g., CPU, RAM, and storage) and running applications that enable the car local resource infrastructure to host MEC-compliant applications. In addition, the \textit{car} module forwards management messages to prepare and configure the local virtualization infrastructure received from the MEC-H management entities. Specifically, the module runs the \textit{ClientResApp}, which requests and decides whether to accept rewards and handles resource registration and release (section~\ref{subsec:interaction}). Moreover, it executes the \textit{VIApp} that applies management instructions received from the MEC-H, to perform local resource allocation and release. It  should be noted that the \textit{VIApp} module can properly carry out its functionalities on any host having a resource infrastructure, thus it might be utilized to allow any device to become part of the MEC-H resource pool.

With the MEC-H now responsible for managing both local and remote resources, the VIM takes on the critical role of scheduling, preparing, and releasing both local and remote resources. To support this functionality, the VIM module implemented in our simulation platform sets up the remote virtual infrastructure (VI) to handle remote commands for allocating, relocating, and terminating MEC applications. Moreover, the VIM supports interchangeable scheduling algorithms to determine the optimal remote host for deploying applications based on the surrounding environment. 

According to our MEC extended architecture \cite{feraudo2023novel}, far-edge devices may leave the resource pool while running MEC applications. When a vehicle hosting MEC applications exits a parking lot in our simulation scenario, it triggers the migration procedure to prevent service interruption and delay. This highlights the importance of designing a reliable migration mechanism for MEC applications in vehicular computing environments. Our framework supplies a module representing an extended version of the Application Mobility Service that transparently addresses resource volatility issues. As previously mentioned in Section~\ref{subsec:interaction}, it supports MEC-assisted intra-host migration. Thus, when an application instance is relocated, e.g., from a remote to a local resource infrastructure, the module requires the MEC application to transfer the user context associated with the NRUE device it is serving.

To summarize, the novel modules included in our simulation platform provide an unprecedented solution for researchers to develop and test MEC-compliant applications in a vehicular computing environment that leverages parked vehicles as sources of virtualized resources, in particular of virtualized processing. Our platform also facilitates the design of novel scheduling algorithms for identifying a suitable host for deploying MEC applications: researchers can evaluate various metrics during their design, such as the average latency between a host and the central infrastructure, or the probability of a node contributing further to the resource pool based on historical data.



\begin{figure}[t!]
    \centering
    \includegraphics[width=1\linewidth]{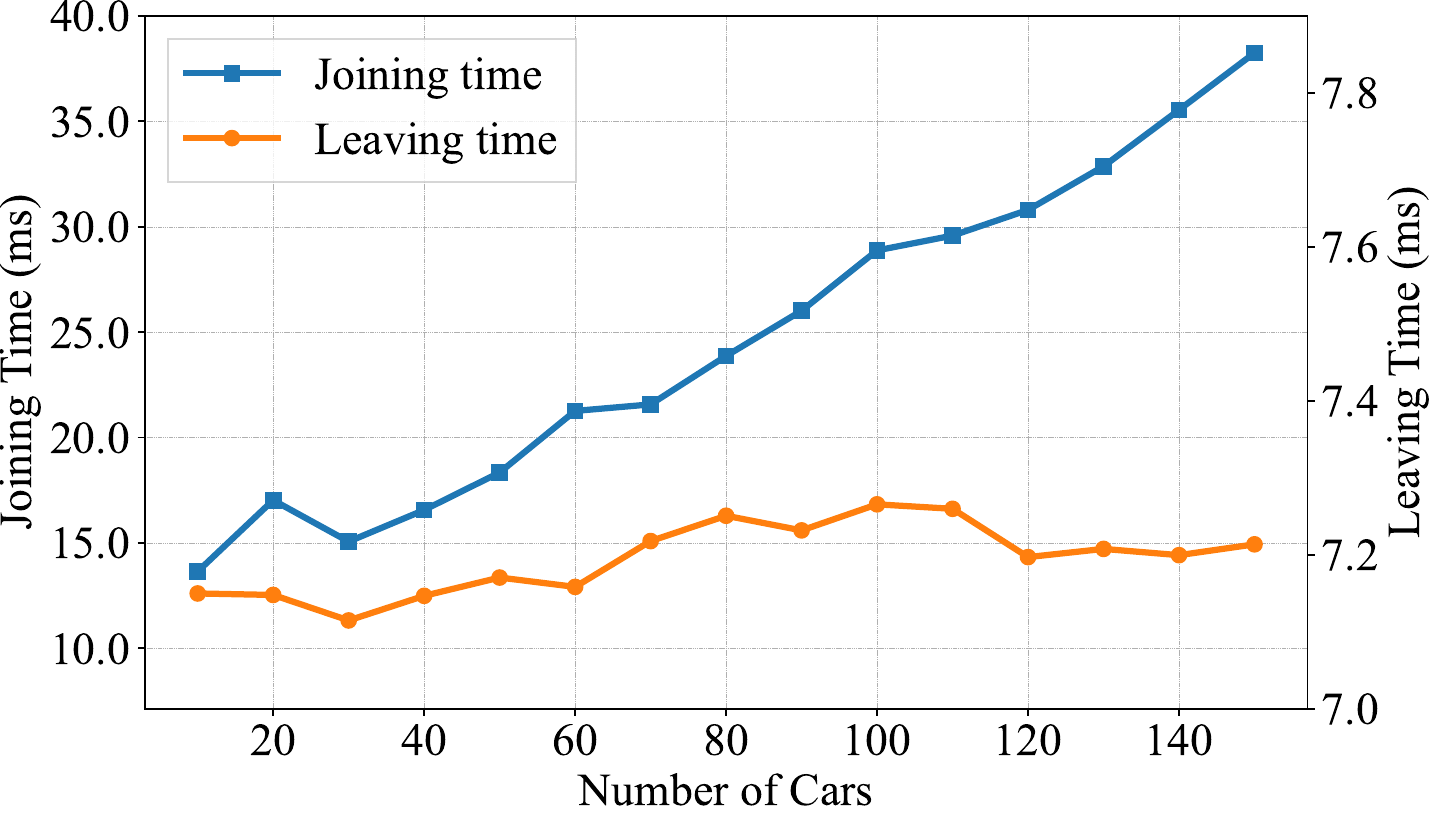}
    \caption{Resource acquisition and releasing protocol times}
    \label{fig:protocols}
\end{figure}

\begin{figure}[t]
    \centering
    \includegraphics[width=1\linewidth]{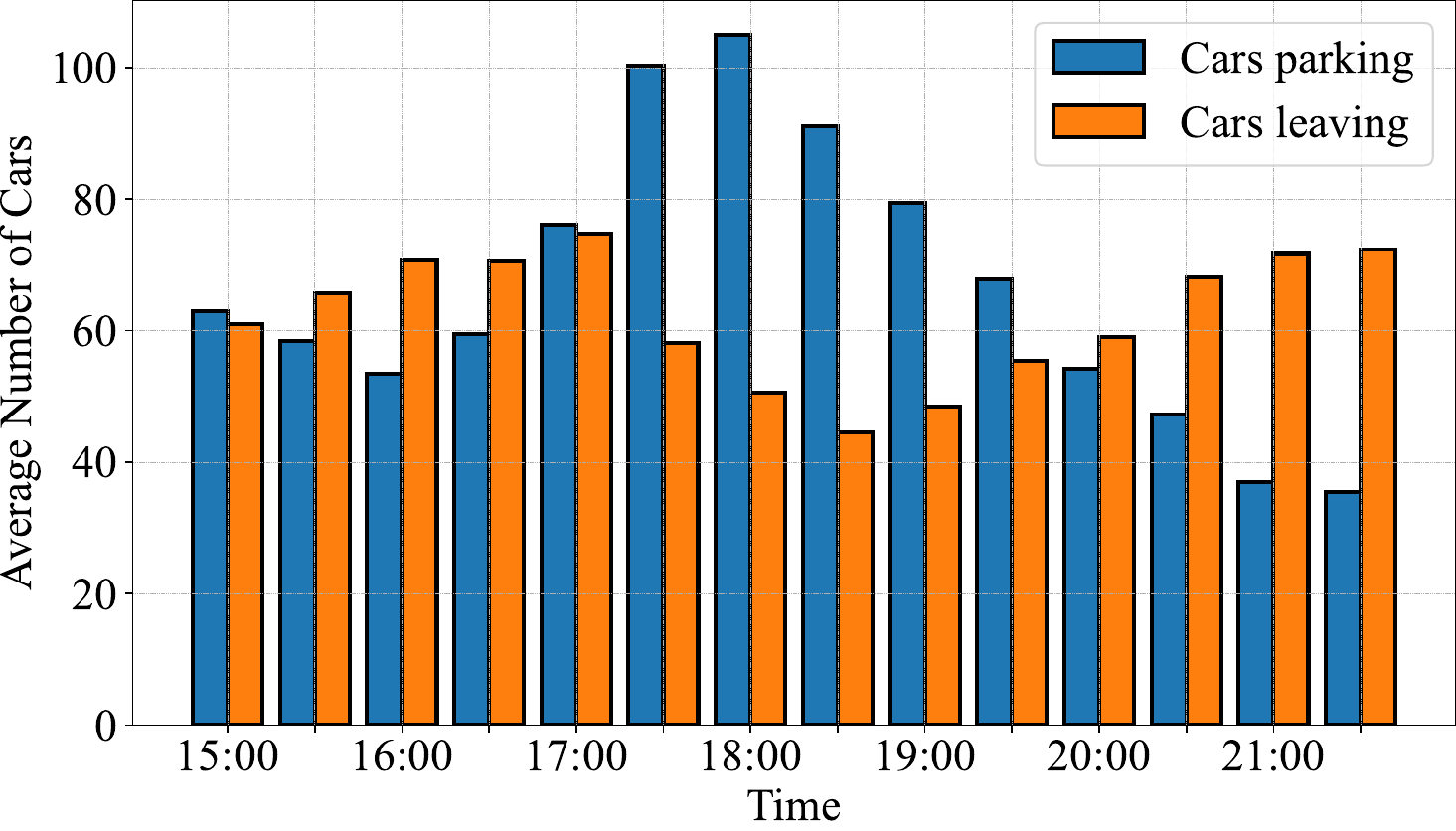}
    \caption{Ceentral Garage City of Arnhem entering and leaving vehicles}
    \label{fig:ceentral}
\end{figure}
\section{\uppercase{Performance Evaluation}}\label{sec:perf}
In this section, we report about some relevant performance indicators measured on top of our simulation platform for some examples of vehicular cloud computing applications. Additionally, we show how our simulation platform can be utilized to define and evaluate an algorithm that efficiently distributes MEC applications on stationary vehicular resources.

Our experiments were conducted in a 5G standalone network environment with a numerology index $\mu=2$. The network consists of a single MEC-H that is connected to a gNB. The scenario involves a parking area located close to the gNB. We implemented a basic reward scheme for the resource acquisition procedure, which utilizes integer values accepted by all participating vehicles. The experiments were carried out on a Linux Virtual Machine running OMNeT++ having 16 CPUs and 64Gb of RAM.
\begin{figure}[t!]
    \centering
    \includegraphics[width=1\linewidth]{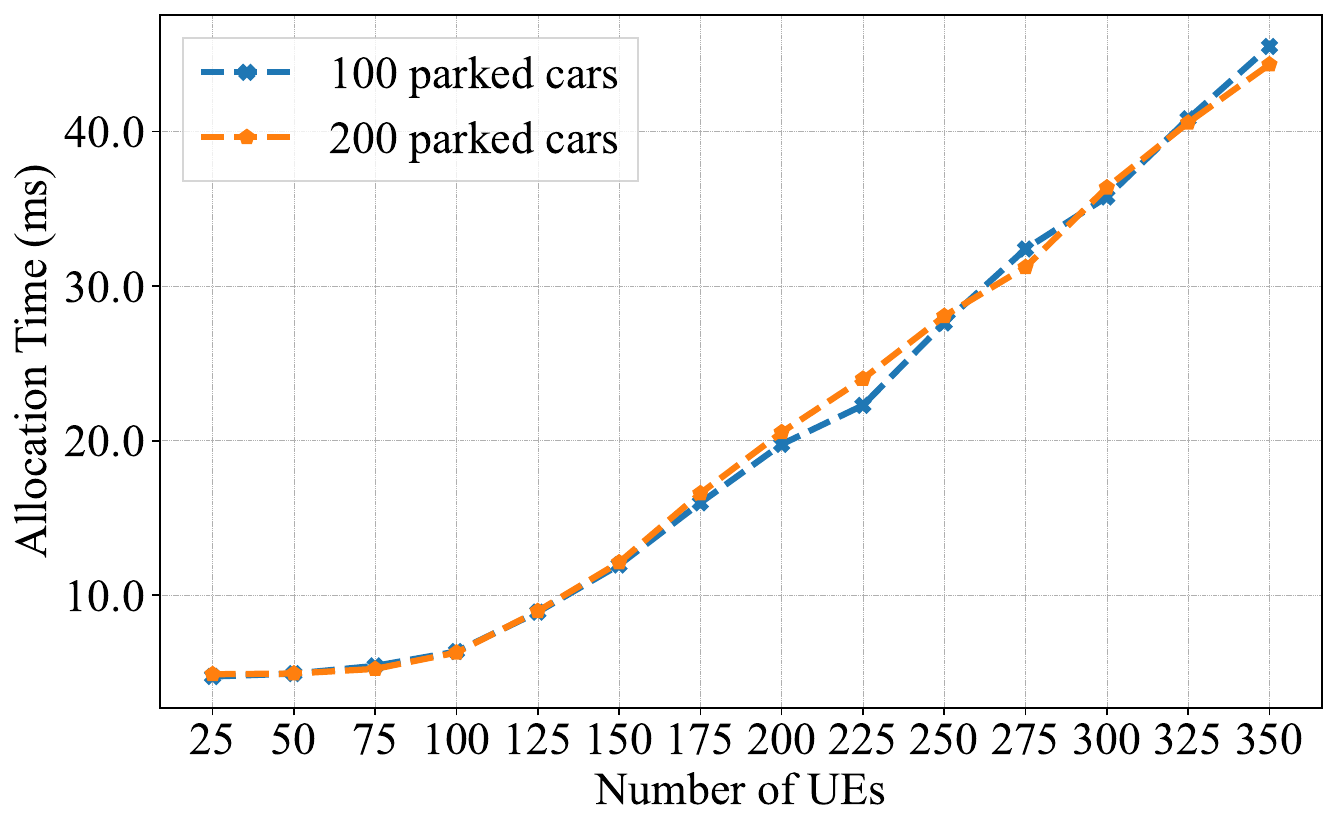}
    \caption{Resource Allocation Time}
    \label{fig:alltime}
\end{figure}
\subsection{Resource Management}
As described in section~\ref{subsec:interaction}, the resource management process involves the operations required to ensure the correct acquisition, allocation, and release of remote resources.

Figure \ref{fig:protocols} illustrates the time required by the protocols for collecting and releasing resources from vehicles as they enter or leave the parking lot within the MEC-H AoI. The join time shown in the figure indicates the time interval for the MEC-H to recognize the availability of a new vehicle for MEC application allocation (step (1)-(6) in Figure \ref{fig:resSub}). On the other hand, the release time is the interval required by the MEC-H to remove the vehicle from the resource pool (steps (1)-(4) in Figure \ref{fig:resRel}). The figure indicates that the join time follows a growing trend ranging from 13 to 40 ms as the number of cars participating in the resource acquisition procedure increases, whereas the release time remains relatively constant (around 7 ms). The difference in performance between the join and release times can be attributed to the varying number of request/response messages generated by the two protocols. On the one hand, the resource release process necessitates only a few messages to exclude a vehicle effectively from the pool. On the other hand, the resource acquisition process involves a series of request/response messages because the device-initiated reward scheme mandates that the vehicle request available rewards. However, it is unlikely for a large number of vehicles to enter a parking lot simultaneously. Such a scenario may only occur during special events like festivals or football matches. To support this claim, we analyzed the data of three parking garages in the city of Arnhem, which is available on the Open Parkeerdata portal\footnote{\url{https://parkeerdata.nl/opendata/arnhem/parkeergarages/transactiedata-parkeergarages}}. Figure \ref{fig:ceentral} depicts the average number of cars entering and leaving the most used garage during rush hours. The figure clearly shows that the number of parked vehicles reaches almost the maximum considered in our test setup between 17:30 and 18:30. Moreover, it is important to note that the peak of participating vehicles does not necessarily occur simultaneously because data were sampled with 30-minute periodicity.

To analyze the time required to allocate MEC applications on remote resources, we measured the delay introduced by the interactions between the VI and VIM during steps (7)-(8) of the process in Fig.~\ref{fig:rra}. The simulation involves multiple UEs requesting MEC app execution and several parked cars belonging to the MEC-H resource pool. MEC applications are evenly distributed on remote nodes using a Round Robin scheduler. We repeated the simulation 10 times with varying numbers of UEs and parked cars.

Figure~\ref{fig:alltime} illustrates that the delays associated with resource allocation follow an exponential growth that depends on the number of MEC applications deployed on parked cars. This is confirmed by the overlapping curves, which indicate that the delay values remain relatively constant even when the number of parked cars varies. It is worth mentioning that we simulated the worst-case scenario, in which all UEs requested MEC app execution simultaneously, thus leading to a substantial increase in network traffic. Despite this, the delay caused by these interactions remained negligible, even when the number of requests exceeded 300, with a delay of approximately 40ms.

\subsection{A Custom Scheduler for Stationary Vehicular Resources}
\begin{figure}[t!]
    \centering
    \includegraphics[width=1\linewidth]{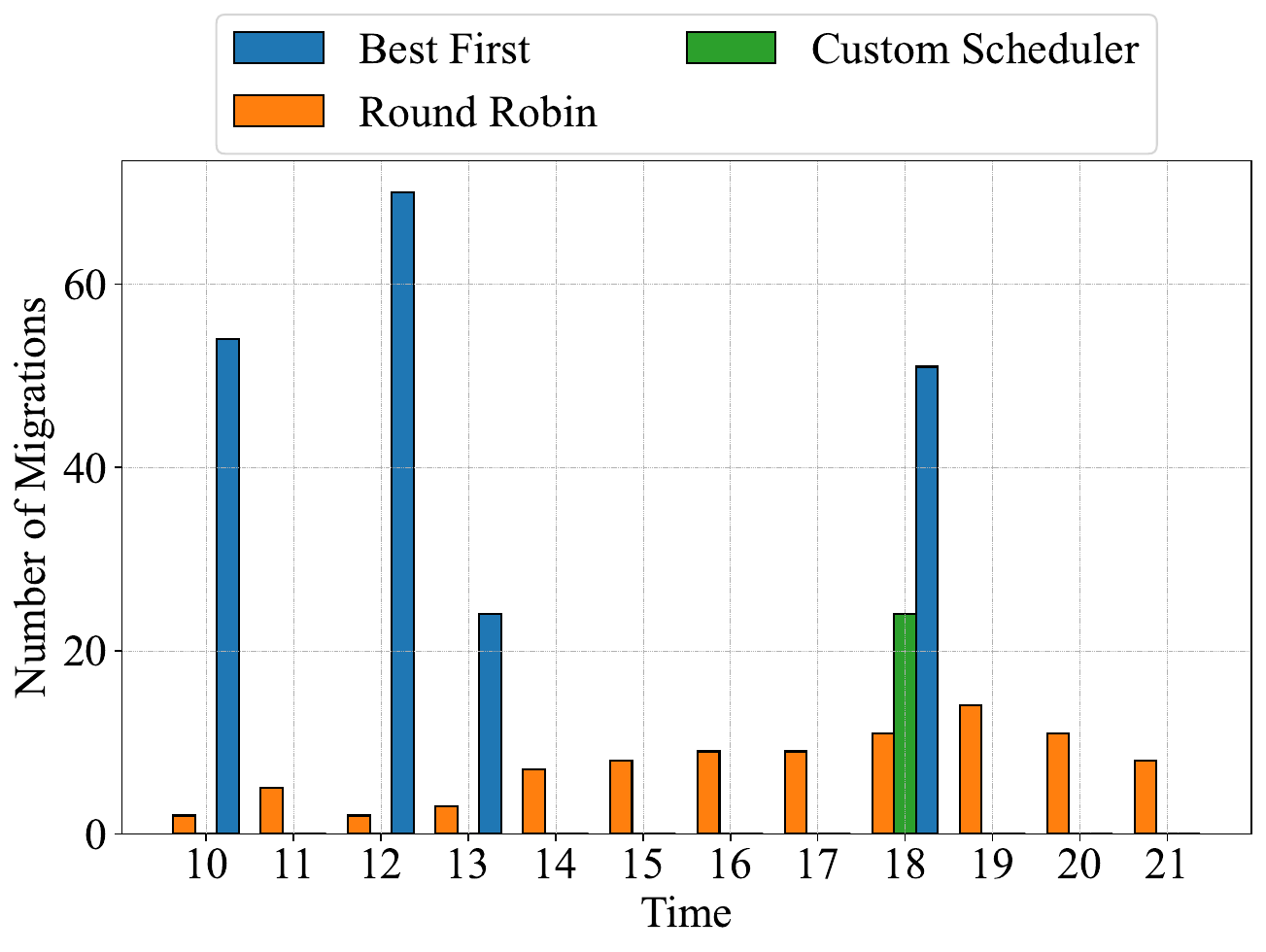}
    \caption{Comparison Scheduling Algorithms in terms of Migrations}
    \label{fig:scheduling}
\end{figure}

To show an example of the usage of our simulation platform to assist researchers in the design, evaluation, and assessment of new algorithms and protocols, we have built and tested the performance of a custom scheduling algorithm that distributes MEC applications on remote resources offered by parked vehicles. It aims at minimizing the number of migrations generated by vehicles leaving the parking lot while running applications. 

To generate vehicle and user behaviors, we recreated the scenario utilized in~\cite{feraudo2023novel}. This approach involves constructing a series of Poisson and Gaussian distributions using two real-world datasets. The Arnhem dataset, already presented in the previous section, was used to model the distributions describing vehicle entry and residency times in a parking lot. The Bologna WiFi dataset\footnote{\url{https://opendata.comune.bologna.it/explore/dataset/iperbole-wifi-affluenza/information/}} provided information on user activities on Open WiFi networks within the city of Bologna, which enabled the creation of distributions mimicking the user behavior during each hour of the day. As in our previous work, we assume that each vehicle that enters the parking lot accepts the rewards proposed by the MEC-H, and each user requests triggers the execution of a one-to-one MEC application.

We considered a simulated period of 24 hours of vehicle and user activities. We run three simulations, one for each scheduling algorithm, namely best first, round robin, and our custom algorithm. The performance of these algorithms was evaluated based on the number of migrations they generated, as this directly impacts the reliability of MEC applications. In other words, a lower number of migrations is desirable for improved performance. 

Figure~\ref{fig:scheduling} reports the associated performance results, by referring to the most challenging case of the day hours with highest levels of user and vehicle activity. The best first algorithm chooses the first available vehicle from the pool that has sufficient resources to execute the application. However, this approach can lead to a large number of migrations, as the selected vehicles may leave the parking lot while running all the applications they are capable of executing. In fact, the number of migrations exceeds 60 at the 12th hour of the simulation. Conversely, the round-robin algorithm maintains a steady number of migrations (around 7.42 in average)as the applications are equally distributed on the vehicles belonging to the MEC-H resource pool.
In addition, by using our simulation platform, we have developed a custom scheduler that relies on multiple Gaussian distributions by using means and standard deviation produced after a pre-processing phase of the Arnhem dataset, which generated the average occupancy time based on a 10-minute interval sampling. Hence, for each vehicle that belongs to the resource pool, the custom scheduler utilizes the time at which it joined the pool and the aforementioned distributions to predict its residency time. It then assigns the MEC application to the vehicle with the highest remaining residency time. The results in the figure demonstrate how this algorithm can largely over-perform the others in the considered application scenario: even if it could be enhanced via more sophisticated machine learning techniques, already in its simple current version it generates only around 20 migrations at the 18th hour of the simulation.


\section{\uppercase{Conclusive Remarks}}\label{sec:concl}
This paper originally presents a simulation platform capable of assisting researchers in the development, evaluation, and assessment of vehicular application algorithms and protocols that exploit vehicular cloud resources accessed according to the standard ETSI MEC specifications. After presenting our extended MEC architecture for these scenarios, the paper reports about the design and implementation of our original simulation platform, with its resource management modules and procedure interactions. In addition, this paper originally describes a concrete example of how our simulation platform can be utilized to design, test, and implement applications that exploit the vehicular computing paradigm.



\bibliographystyle{apalike}
{\small
\bibliography{reference}}



\end{document}